\documentclass[a4paper,11pt]{article}
\pdfoutput=1
\usepackage{jheppub} 
\usepackage[T1]{fontenc}
\usepackage{booktabs} 
\usepackage{color}
\usepackage{subcaption}
\usepackage{multirow}      
\usepackage{array}  
\newcounter{asteriskfootnote}
\newcommand*{\asteriskfootnote}[1]{%
\setcounter{asteriskfootnote}{\value{footnote}}%
    \renewcommand*{\thefootnote}{\fnsymbol{footnote}}%
    \footnote[1]{#1}%
    \setcounter{footnote}{\value{asteriskfootnote}}%
    \renewcommand*{\thefootnote}{\arabic{footnote}}%
    }
   
   \newcounter{daggerfootnote}
\newcommand*{\daggerfootnote}[1]{%
\setcounter{daggerfootnote}{\value{footnote}}%
 \renewcommand*{\thefootnote}{\fnsymbol{footnote}}%
  \footnote[2]{#1}%
   \setcounter{footnote}{\value{daggerfootnote}}%
   \renewcommand*{\thefootnote}{\arabic{footnote}}%
}

\title{\huge Black hole interiors of  homogeneous holographic solids under shear strain}

\author[a]{Yuanceng Xu,}
\author[b,c,d]{Li Li,\asteriskfootnote{Corresponding author}}
\author[a]{and Wei-Jia Li\daggerfootnote{Corresponding author}}
\affiliation[a]{Institute of Theoretical Physics, School of Physics,
 Dalian University of Technology, Dalian 116024, China}
\affiliation[b]{Institute of Theoretical Physics, Chinese Academy of Sciences, Beijing 100190, China}
\affiliation[c]{School of Physical Sciences, University of Chinese Academy of Sciences, Beijing 100049, China}
\affiliation[d]{School of Fundamental Physics and Mathematical Sciences,
 Hangzhou Institute for Advanced Study, UCAS, Hangzhou 310024, China}
\emailAdd{xuyc@dlut.edu.cn} 
\emailAdd{liliphy@itp.ac.cn}
\emailAdd{weijiali@dlut.edu.cn} 

\abstract{We investigate the interior of AdS black holes under finite shear strain in a class of holographic axion models, which are widely used to describe strongly-coupled systems with broken translations. We demonstrate that the shear anisotropy necessarily eliminates the inner Cauchy horizon, such that the deformed black hole approaches a spacelike singularity. 
The anisotropic effect induced by the axion fields 
triggers a collapse of the Einstein-Rosen bridge at the would-be Cauchy horizon, accompanied by
a rapid change in the anisotropy of the spatial geometry. In addition, for a power-law axion potential, sufficiently large shear deformations give rise to a domain wall solution that includes a Lifshitz like scaling geometry near the boundary as well as a near horizon Kasner epoch with the Kasner exponents determined by the powers of the potential. Finally, we find that the interior dynamics of black holes generally enter an era described by an anisotropic Kasner universe at later interior time. Depending on the form of the potential, they either tend to stable Kasner universes, or exhibit an endless alternation of different Kasner epochs toward the singularity.}

\begin{document} 
\maketitle

\section{Introduction}
\label{sec:intro}

Since the establishment of general relativity, understanding the nature of black holes has become a significant scientific question. In recent years, observational breakthroughs, including gravitational-wave detection~\cite{LIGOScientific:2016aoc} and direct imaging of black hole shadows~\cite{EventHorizonTelescope:2019dse,EventHorizonTelescope:2019ths}, have further enriched our empirical knowledge of black hole exteriors. However, the physics of black hole interiors remains far less understood, primarily due to the presence of a curvature singularity enclosed by the event horizon, where conventional physical laws cease to hold. The inevitability of such singularities within general relativity was rigorously established by Penrose and Hawking~\cite{Penrose:1964wq,Hawking:1970zqf}. In contrast to the spacelike singularity of the Schwarzschild solution, the singularity of a Reissner-Nordström (RN) black hole or a Kerr black hole is timelike owing to the presence of a Cauchy horizon. Moreover, the strong cosmic censorship (SCC) conjecture~\cite{Isenberg:2015rqa} 
asserts that any Cauchy horizon must be dynamically unstable to preserve the causal integrity of the spacetime~\cite{Ringstrom:2015jza}.

A seminal contribution to the understanding of black hole interiors is the Belinski-Khalatnikov-Lifshitz (BKL) conjecture~\cite{Lifshitz:1963ps,Belinsky:1970ew,Belinski:1973zz} which describes the dynamical behavior of spacetime near a spacelike singularity. According to this, the geometry undergoes chaotic, oscillatory dynamics characterized by an infinite sequence of Kasner epochs—periods of anisotropic expansion or contraction governed by distinct Kasner exponents. For a large number of models, the alternation between these epochs can be elegantly described using a cosmological billiard picture~\cite{Damour:2002et,Henneaux:2007ej}, wherein the system’s evolution is mapped to a billiard ball reflecting off the walls of a multi-dimensional potential. This analogy provides a powerful tool for analyzing the chaotic nature of the BKL singularity and has profound implications for both the physics of black hole interiors and the early-universe cosmology.

The study of black hole interiors is further motivated by holographic duality—a profound theoretical link between weakly coupled gravitational systems and strongly coupled quantum field theories. As a cross‑disciplinary framework, it also provides novel pathways for probing the interior structure of black holes. Some dual field theory quantities, in particular, correlation functions~\cite{Fidkowski:2003nf,Festuccia:2005pi}, entanglement entropy ~\cite{Hartman:2013qma} and quantum complexity~\cite{Stanford:2014jda,Brown:2015bva} serve as complementary non-perturbative probes of interior dynamics.

Recent studies have revealed that coupling matter fields to AdS-Schwarzschild~\cite{Frenkel:2020ysx,Hartnoll:2020rwq} or RN black holes~\cite{Hartnoll:2020fhc} can dramatically modify their interior dynamics, particularly through the elimination of Cauchy horizons~\cite{Cai:2020wrp,An:2021plu,Hale:2025ezt}. The resulting dynamical evolution typically progresses through three characteristic phases: collapse of Einstein-Rosen (ER) bridge, Josephson oscillations and Kasner behavior near the singularity~\cite{Hartnoll:2020fhc} - with the latter potentially exhibiting Kasner inversions, transitions and reflections~\cite{Cai:2023igv,Cai:2024ltu}. These phenomena have been verified in a broad class of systems within the framework of holography, including holographic superconductors~\cite{Hartnoll:2020fhc,Sword:2022oyg,Cai:2021obq,Cai:2024ltu,Zhang:2025tsa,Zhang:2025hkb}, holographic metals~\cite{Gao:2023zbd,Carballo:2024hem}, holographic axion models~\cite{Mansoori:2021wxf,Mirjalali:2022wrg,Prihadi:2025rwg,Sword:2021pfm,Arean:2024pzo} as well as the modified gravity with higher-derivative corrections~\cite{Devecioglu:2021xug,Devecioglu:2023hmn,Grandi:2021ajl,Caceres:2024edr,Bueno:2024fzg}. A common conclusion from these investigations is the necessity of coupling additional matter fields to the original gravitational theory to produce richer interior structures.

Building on these developments, we examine black hole interiors within anisotropic holographic axion models~\cite{Baggioli:2021xuv},
with two crucial distinctions from previous work~\cite{Mansoori:2021wxf,Mirjalali:2022wrg,Prihadi:2025rwg,Sword:2021pfm,Arean:2024pzo}: (1) our construction utilizes only the axion fields without additional (charged) scalar hair; and (2) the bulk axion fields realize the anisotropy on the boundary in a specific manner that allows finite shear deformations to be introduced in the holographic description\footnote{Previous studies focused on isotropic cases in the presence of axions, for which a proof for no inner-horizon is absent~\cite{Mirjalali:2022wrg,Prihadi:2025rwg,Sword:2021pfm}.}. This model has been applied to study nonlinear elasticity and mechanical failure of amorphous solids~\cite{Baggioli:2020qdg,Pan:2021cux,Ji:2022ovs,Baggioli:2023dfj}, and more recently, it uncovers an unexpected connection between non-linear elasticity and quantum complexity~\cite{Xu:2025uzo}. However, in this work, we focus instead on its implications for black hole interiors - particularly how shear deformations, crucial to analyzing the elasticity of boundary systems, manifest in interior spacetime dynamics.  

We first prove that shear anisotropy necessarily eliminates the inner Cauchy horizon, such that the deformed black hole geometry terminates at a spacelike singularity in contrast to numerous cases of holographic axion models where isotropic black holes, in general, have inner Cauchy horizons and a timelike singularity. 
We then focus on the dynamics of the black hole interior. The instability of the inner Cauchy horizon triggered by the small shear anisotropy leads to a rapid collapse of the ER bridge at the would-be inner horizon, together with a rapid change in the anisotropy of the spatial geometry.
For sufficiently large shear deformation and the model with a power-law potential, the occurrence of ER bridge collapse is replaced by the emergence of an intermediate domain wall geometry spanning the event horizon. It displays an anisotropic Lifshitz like scaling near the boundary, and satisfies a general Kasner form near the singularity whenever the potential term is subdominant. On the contrary, when the potential term becomes important(\emph{e.g.} exponential like potentials),  the interior geometry can never reach an asymptotic scaling behavior. Instead, the evolution of the black hole interior exhibits an infinite sequence of Kasner epochs.

This paper is organized as follows: Section~\ref{sec:setup} reviews the holographic axion model incorporating finite shear strain.
In Section~\ref{sec:nohorizon}, we rigorously demonstrate that anisotropic shear deformations induce the disappearance of the inner horizon. Section~\ref{sec:nearhorizon} presents numerical and analytical evidences for both the collapse dynamics of the ER bridge under small shear strains and a domain wall geometry induced by large shear deformations.  The influence of shear deformation on near-singularity Kasner epoch is systematically investigated in Section~\ref{sec:kasner}.
Finally, we conclude in Section~\ref{sec:conclusion} and propose several directions for future research. More technical details are provided in Appendix~\ref{appa1} for the constraint on the axion potential and in Appendix~\ref{appa} for the analytic analysis of the domain wall geometry under large shear deformations.

\section{Holographic model}
\label{sec:setup}

We consider the general holographic axion model as follows~\cite{Baggioli:2014roa,Alberte:2015isw},
\begin{eqnarray}
    S=\int d^{4}x \sqrt{-g}\left[R-2\Lambda-2m^2V(X,Z)   \right],\label{action}
\end{eqnarray}
with
\begin{eqnarray}\label{eqIJ}
\mathcal{I}^{IJ}\equiv\partial_{\mu}\phi^{I}\partial^{\mu}\phi^{J},\  \  X\equiv\frac{1}{2}\text{Tr}\left[\mathcal{I}^{IJ}\right],\  \  Z\equiv\det[\mathcal{I}^{IJ}],\   \  I,J=\{1,2\},
\end{eqnarray}
where $R$ is the Ricci scalar, $\Lambda$ is the cosmological constant that will be fixed as $\Lambda=-3$ so that the AdS radius is normalized, $m^2$ is an effective coupling with the dimension of mass squared, the potential $V(X,Z)$ is a general function of $X$ and $Z$. 
The self-consistency of the bulk theory combined with the requirement for the well-defined elastic property of the boundary system, such as a positive shear modulus, necessitates that $V_X\geq0$ (see Appendix~\ref{appa1})\footnote{ This condition is crucial for the proof in Section~\ref{sec:nohorizon}  that shear deformations  lead to the disappearance of the inner horizon universally.}.
In addition, the massless scalars $\phi^I(t,u,x^i)$ are usually referred to as `axions' in some literatures because of the internal shift symmetry. In this work, we concentrate only on the $3+1$-dimensional bulk spacetime. However, the construction can be easily generalized into higher dimensions \cite{Xu:2025uzo}. To obtain a homogeneous background that is independent of the spatial coordinates, $x$ and $y$, one can consider the bulk solution with a scalar hair\footnote{Throughout the paper, we denote background values of the quantities with bars.}
\begin{eqnarray}   
\bar{\phi}^{I}={M^{I}}_{i}x^{i},\quad   i=\{1,2\},  \label{axions}
\end{eqnarray}
where ${M^I}_i$ is a $2\times2$ symmetric matrix, the notations $x^1\equiv x$ and $x^2\equiv y$ have been adopted. This class of models is nothing but the St\"uckelburg formalism
of the Lorentz-violating massive gravity dual to broken translations on boundary \cite{Vegh:2013sk}. 

From the boundary perspective, the profile of axions $\phi^I\sim x^i$ cannot determine whether the breaking of translations is spontaneous or explicit. Instead, this should be fixed by the specific choice of the potential $V$. Sticking to the standard quantization, if $V$ decays much faster than $u^5$ near the boundary ($u=0$), there exist gapless phonon excitations with the speeds correlated with the elastic response of the system which corresponds to the spontaneous symmetry breaking (SSB) of translations \cite{Alberte:2017oqx}.\footnote{One can have a well-defined elastic response for potentials that decay at the boundary as $u^3$ or faster.} In this case, the gravity system is dual to a homogeneous solid on boundary. On the contrary, if $V$ decays slower than $u^5$ near the boundary, the symmetry breaking is explicit which 
leads to momentum relaxation \cite{Andrade:2013gsa}. For the relationship between the mechanism of translational symmetry breaking and the near-boundary behavior of $V$, one may refer to Appendix \ref{appa1} for further details.
The most part of this work will focus on the former case. However, an extended discussion about the later case will be attached at the end.

In the purely SSB pattern, the profile of the axions $\bar{\phi}^I\sim x^i$ should be interpreted as the vacuum expectation value (vev) of scalar operators $\Phi^I(t,x^i)$ on boundary.
These scalar operators act
as a set of co-moving coordinates in the Lagrangian representation of solids and their vev selects a specific configuration. For the system in equilibrium, the configuration can be chosen as $\langle\Phi^I\rangle=\delta^I_ix^i=\bar{\phi}^I$   \cite{Nicolis:2013lma}. On top of this, certain mechanical deformation can be introduced by parameterizing the matrix ${M^{I}}_{i}$ properly. When we set \cite{Alberte:2018doe}
\begin{eqnarray}   
 {M^{I}}_{i}=\begin{pmatrix}
    \sqrt{1+\epsilon^2/4} &  \epsilon/2 \\
    \epsilon/2             &   \sqrt{1+\epsilon^2/4}
\end{pmatrix},  \label{shear}
\end{eqnarray}
with a dimensionless variable $\epsilon$, the deformed configuration further breaks the rotational symmetry in the $x-y$ plane. The constraint $\text{det}\,{({M^I}_i)}=1$ implies that the deformation corresponds to a purely shear strain which preserves the spatial area of the 2+1-dimensional boundary system.\footnote{In cases of explicit symmetry breaking (ESB), $\epsilon$ no longer describes a shear strain but merely quantifies the strength of anisotropy.}

In the presence of a non-zero $\epsilon$, the metric ansatz for a homogeneous (planar) black hole takes the following form
\begin{eqnarray}
ds^2=\frac{1}{u^2}\left[-f(u)e^{-\chi(u)}dt^2+\frac{1}{f(u)}du^2+\gamma_{ij}(u)dx^{i}dx^j\right],  \label{metic}
\end{eqnarray}
with
\begin{eqnarray}
    \gamma_{ij}(u)=\begin{pmatrix}
        \cosh[h(u)]  &\  \  \sinh[h(u)] \\ 
        \sinh[h(u)]  &\  \  \cosh[h(u)]
    \end{pmatrix},
\end{eqnarray}
which ensures the compatibility with \eqref{shear} and $\text{det}(\gamma_{ij})=1$ for any $h(u)\geq0$. Similarly, one can reparameterize the deformation matrix as follows
\begin{eqnarray}
     {M^{I}}_{i}=\begin{pmatrix}
    \cosh(\Omega/2)  &\   \  \sinh(\Omega/2) \\
     \sinh(\Omega/2) &\   \   \cosh(\Omega/2)
\end{pmatrix},
\end{eqnarray}
 which implies that the shear strain can be read via the relation $\epsilon=2\sinh\left(\Omega/2\right)$.
 Then, the background equations are given by 
\begin{eqnarray}
\chi'(u)-\frac{1}{2}uh'(u)^2&=&0,  \label{EOM3}\\
\frac{f'(u)}{f(u)}-\frac{3}{u}-\frac{\chi'(u)}{2}+\frac{3-m^2V(\bar{X},\bar{Z})}{u f(u)}&=&0,  \label{EOM4}\\
h''(u)+h'(u)\left[\frac{f'(u)}{f(u)}-\frac{2}{u}-\frac{\chi'(u)}{2}\right]+\frac{2m^2\sinh{[\Omega-h(u)]}V_X(\bar{X},\bar{Z})}{f(u)}&=&0,  \label{EOM5}
\end{eqnarray}
where $\bar{X}=u^2\cosh[\Omega-h(u)]$ and $\bar{Z}=u^4$.
The Hawking temperature and the entropy density can be obtained from
\begin{eqnarray}\label{axionRN}
    T=-\frac{f'(u_h)e^{-\chi(u_h)/2}}{4\pi}=\frac{3-m^2V(\bar{X}_h,\bar{Z}_h)}{4\pi u_h}e^{-\chi(u_h)/2},\quad  s=\frac{4\pi}{u^2_h}\,, \label{therm}
\end{eqnarray}
with $u_h$ denoting the location of the event horizon that satisfies $f(u_h)=0$. Here, we have introduced $\bar{X}_h\equiv\bar{X}(u_h)$ and $\bar{Z}_h\equiv\bar{Z}(u_h)$.

In the isotropic case with no shear deformation $(\epsilon=0)$, the metric functions $f(u)$, $\chi(u)$ and $h(u)$ can be solved analytically, yielding

\begin{eqnarray}
    f(u)=u^3\int^{u_h}_{u}\frac{3-m^2V(s^2,s^4)}{s^4}ds,\    \   \chi(u)=h(u)=0\,. \label{eq211} 
\end{eqnarray}
The inclusion of the axion term permits the existence of two horizons satisfying $f(u)=0$ in Eq.~\eqref{eq211}.
Their locations are denoted as $u_h$ and  $u_i$, respectively (We have $u_h<u_i$ in the present coordinates), where $u_i$ we refer to as the axion inner horizon\footnote{The precise location of $u_i$ is determined by by the integral condition: $\int^{u_h}_{u_i}\left(\frac{3}{s^4}-\frac{m^2V(s^2,s^4)}{s^4}\right)ds=0$ according to Eq.~\eqref{eq211}. This condition admits a broad range of potential forms $V$, including polynomial functions and a potential that diverges exponentially or even worse. Conversely, no inner horizon exists when $V$ is negative definite.}.

For the finite shear strain, we need solve \eqref{EOM3}-\eqref{EOM5} numerically.  The boundary conditions are specified as follows: At the event horizon $(u=u_h)$, we fix $f(u_h)=0$, $\chi(u_h)=\chi_0$ and $h(u_h)=h_0$, and near the UV boundary $(u\rightarrow0)$, the asymptotic AdS requires that $f(0)\rightarrow1$, $\chi(0)\rightarrow0$, while $h(u)$ behaves as
\begin{eqnarray}
h(u)\simeq \mathcal{H}_0+\dots+\mathcal{H}_3u^3
+\dots.
\end{eqnarray}
According to the holographic dictionary, the leading behavior $\mathcal{H}_0$ corresponds to the external source for the boundary stress tensor $T_{xy}$.
When fixing $\mathcal{H}_0=0$, the coefficient $\mathcal{H}_3$ is proportional to the vacuum expectation value of the stress tensor, \emph{i.e.} the shear stress 
$\sigma\equiv\langle T_{xy}\rangle=3\mathcal{H}_3$ \cite{Baggioli:2020qdg}. With this setup, the shear stress is induced solely by the axion-mediated strain. 

\begin{figure}[htbp]
    \centering
    \includegraphics[width=0.5\linewidth]{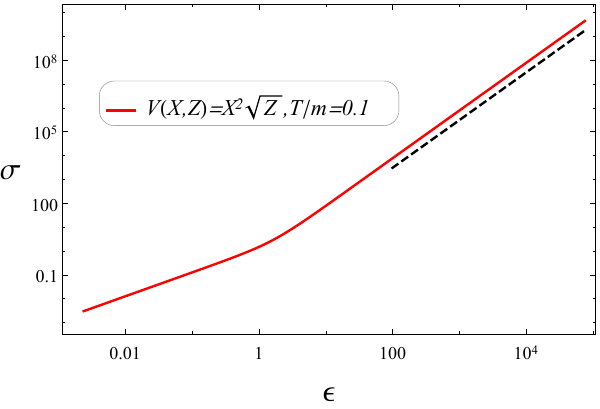}
    \caption{The nonlinear elastic stress-strain curve for the potential $V(X,Z)=X^2\sqrt{Z}$. For sufficiently large shear deformation $\epsilon$, the nonlinear stress-strain curve follows a scaling law of $\sigma\sim\epsilon^{3M/(M+2N)}$. 
    Here we set $T/m=0.1$.}
    \label{stressstraincurve}
\end{figure}

In the following discussion, we will primarily consider the benchmark models with the potential $V(X, Z) = X^MZ^N$. Then, the SSB requires that $M + 2N > 5/2$. The stress-strain curves can be obtained by numerically solving the background equations and applying the holographic dictionary. In Fig.~\ref{stressstraincurve}, we show $\sigma$ as the function of $\epsilon$ for a specific model with $M=2$ and $N=1/2$. A nonlinear power-law behavior $ \sigma\sim\epsilon^{\nu}$ with $\nu=3M/(M+2N)$ in the large strain regime was firstly revealed in ~\cite{Baggioli:2020qdg}.

\section{Shear deformations remove the Cauchy horizon}\label{sec:nohorizon}

Next, we will show that the shear deformation dramatically alters the interior structure of black holes. It is noted that previous methods for proving the non-existence of a second horizon, such as those relying on a conserved charge~\cite{Cai:2020wrp,Hartnoll:2020fhc} or the null energy condition~\cite{An:2021plu}, fail due to the axion term. Nevertheless, for black holes under shear strain, we can still prove the absence of an inner horizon.

Suppose that there exist two horizons: an event horizon located at $u_h$ and an inner horizon at $u_i$, such that 
\begin{eqnarray}
    f(u_h)=f(u_i)=0,\  \  \text{with} \   \  u_h<u_i.
\end{eqnarray}
Between these two horizons, the blackening factor satisfies $f(u)<0$. For the black hole under shear strain (with nonzero $h$ and $\Omega$), the following equation can be derived from Eq.~\eqref{EOM5}:
\begin{eqnarray}
   \left[ \frac{e^{-\chi(u)/2}f(u)h'(u)}{u^2}\right]'\frac{1}{\sinh[\Omega-h(u)]}+\left(\frac{e^{-\chi(u)/2}}{u^4}\right)2m^2u^2V_X=0,  \label{eom6}
\end{eqnarray}
which implies that 
\begin{eqnarray}
0&=&\int^{u_i}_{u_h}\left[\frac{e^{-\chi(u)/2}f(u)h'(u)}{u^2\sinh[\Omega-h(u)]}\right]'du\nonumber\\
    &=&\int^{u_i}_{u_h}\frac{e^{-\chi(u)/2}}{u^4}\left\{ -2m^2u^2V_X+u^2f(u)h'(u)^2\coth[\Omega-h(u)]\text{csch}[\Omega-h(u)]\right\}du.  \label{43}
\end{eqnarray}
The first equality follows from the condition $f(u_h)=f(u_i)=0$.  Note that the product of the two hyperbolic functions in the last line is always positive, i.e.,
\begin{eqnarray}
    \coth[\Omega-h(u)]\text{csch}[\Omega-h(u)]> 0\,.
\end{eqnarray}
Therefore, for the potential satisfying $V_X\ge0$  as required by theoretical consistency (see Appendix~\ref{appa1}), the integrand in Eq.~\eqref{43} remains negative throughout the integration domain. According to Eq.~\eqref{eq211}, the existence of two distinct horizons requires the isotropic limit with no shear deformation. This demonstrates that in the general case, anisotropic shear strain necessarily eliminates the inner horizon.

\begin{figure}[htbp]
\centering
\includegraphics[width=0.55\textwidth]{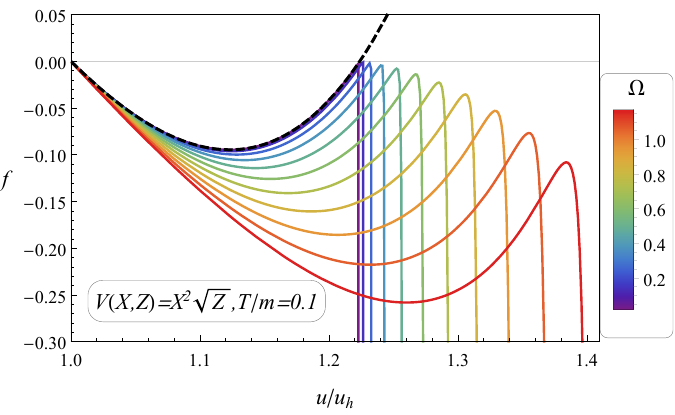}
\hfill
\caption{The blackening factor $f(u)$ near the horizon and inside the black hole for different values of $\Omega$. The black dashed line represents the isotropic solution of $f(u)$  under zero shear strain $(\epsilon=0)$ . Here we set $V(X,Z)=X^2\sqrt{Z}$, $T/m=0.1$. }\label{blacking}
\end{figure}

Fig.~\ref{blacking} shows the blackening factor 
$f(u)$ as a function of $u/u_h$ for various values of $\Omega$, for the model 
$V=X^2\sqrt{Z}$ at $T/m=0.1$. In the isotropic case, the function $f(u)$ vanishes at the inner horizon $u=u_i$ (see the black dashed line in Fig.~\ref{blacking}). However, $f(u)$ becomes negative definite across the entire region inside the event horizon after turning on the shear strain. This clearly demonstrates that any non-zero shear deformation universally removes the inner horizon. Moreover, one can find that $f(u)$ exponentially decays after $u_i$ for small shear deformations.

\section{Interior dynamics near the horizon}\label{sec:nearhorizon}
When the inner horizon is absent, the black hole interior ends at a spacelike singularity as $u\rightarrow \infty$. 
The asymptotic near-singularity behavior will be discussed in the next section. In this section, we focus on a crossover occurring at $u_i$, the location of the would-be inner horizon. We will demonstrate the emergence of a domain wall geometry across the event horizon induced by a sufficiently large shear strain.

\subsection{Collapse of Einstein-Rosen bridge under small shear deformations}\label{sec:collapse}

A number of works have shown that the disappearance of the inner horizon is associated with the rapid decrease in the metric component $g_{tt}$ (proper distance) over a short time interval (along the $u-$direction inside the black hole) near the would-be inner horizon—a dynamical process known as ER bridge collapse. However, it has been demonstrated that this collapse can be suppressed in certain holographic superconductor models with strong Maxwell-scalar coupling parameter~\cite{Sword:2021pfm}.
In contrast to these previous studies, the inner horizon in our theory is removed by pure shear deformations. 
It is therefore interesting to investigate whether an ER bridge collapse occurs in the present case.

\begin{figure}[htbp]
    \centering
    \begin{subfigure}[b]{0.495\textwidth}
        \centering
\includegraphics[width=1\textwidth]{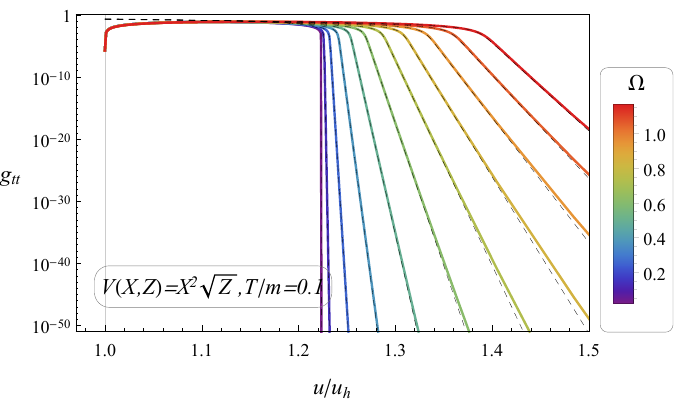}
      \caption{\textbf{(a)}}
    \end{subfigure}
     \hfill   
     \begin{subfigure}[b]{0.495\textwidth}
        \centering
\includegraphics[width=1\textwidth]{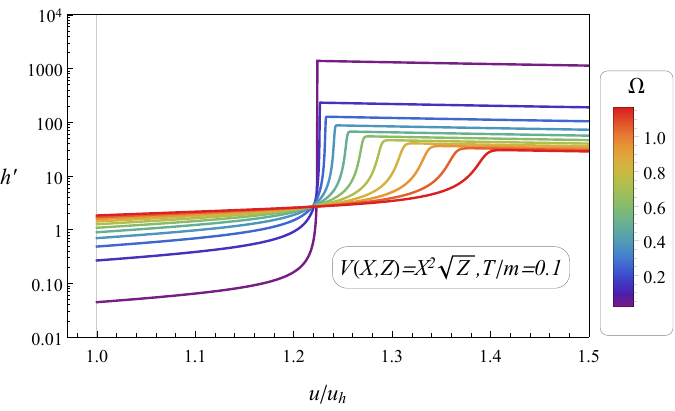}
      \caption{\textbf{(b)}}
    \end{subfigure}
    \caption{\textbf{(a)}: The behavior of the metric component $g_{tt}$ near the would-be inner horizon under different values of $\Omega$. The black dashed curves show semi-analytical results obtained from Eq.~\eqref{eq43}. \textbf{(b)}: 
    The behavior of the function $h'(u)$ with respect to the radial coordinate $u/u_h$ inside the black hole. According to Table.~\ref{tab1}, the maximum value of $h'$ near the would-be inner horizon $u_i$ corresponds to the ratio $c_2/c_1$. We have chosen $V(X,Z)=X^2\sqrt{Z}$ at $T/m=0.1$.}\label{ERcollapse}
\end{figure}

Our proof in Section~\ref{sec:nohorizon} indicates that an arbitrarily small shear deformation results in an instability of the inner horizon and destroys it.
As shown in Fig.~\ref{blacking}, the smaller the shear deformation, the more pronounced the instability. In Fig.~\ref{ERcollapse}, we show the dynamical epoch around the would-be inner horizon due to the instability triggered by the shear deformation.
A typical feature is that the metric component $g_{tt}$ rapidly approaches zero (though not exactly zero) within an extremely short range of  $u/u_h$ near the would-be inner horizon. This process is referred to as the collapse of the ER bridge. Meanwhile, the function $h(u)$, which characterizes the anisotropy of the geometry, undergoes a sudden change near the would-be inner horizon. As shown in Fig.~\ref{ERcollapse} \textbf{(b)}, $h'(u)$  displays a rapid increase and saturates at a finite magnitude.

In addition to the numerical approach, these highly nonlinear spacetime dynamics can be understood as follows. 
For small shear strains, the collapse of the ER bridge occurs very close to the would-be inner horizon. Thus, following the spirit of~\cite{Hartnoll:2020rwq}, we may express $f$,$\chi$ and $h$ as functions of $u=u_i+\delta u$. In the equations of motion, the coordinate $u$ that appears explicitly can be consistently set to $u_i$. 
Numerical verification (or, a posteriori, the solution below) shows that the last term in Eq.~\eqref{EOM5} becomes negligible near the would-be inner horizon because $\Omega-h(u)\approx0$ for small shear strains. With this approximation, the equations of motion~\eqref{EOM3}-\eqref{EOM5}
reduce to the following form:
\begin{eqnarray}
2\chi'&=&u_i(h')^2,  \label{eom1}\\
4u_if'&=&u^2_if(h')^2+4m^2V(u^2_i,u^4_i)-12,  \label{eom2}\\
\left(e^{-\chi/2}fh'\right)'&=&0.  \label{eom3}
\end{eqnarray}
To solve the equations, we begin by integrating Eq.~\eqref{eom3} and writing\\
$h'=-c_1\left[4m^2V(u^2_i,u^4_i)-12\right]^{1/2}e^{\chi/2}/f$, where $c_1$ is a constant. The solution can be expressed in terms of the metric component $g_{tt}=-fe^{-\chi}/u_i^2$, which satisfies 
\begin{eqnarray}
    \frac{g''_{tt}}{g'_{tt}}=\frac{c^2_1g'_{tt}}{g_{tt}(c^2_1+g_{tt})}\,.  \label{eq42}
\end{eqnarray}
This equation admits a general solution of the form (keeping in mind that $g_{tt}>0$ inside the black hole)
\begin{eqnarray}
    c^2_1\log(g_{tt})+g_{tt}=-\frac{u_i}{4}c^2_2(\delta u+c_3)\,,  \label{eq43}
\end{eqnarray}
where $c_2$ and $c_3$ are integration constants (normalized for later convenience). In addition to relations $f=-e^{\chi}g_{tt}u^2_i$ and
$c^2_1g'_{tt}/g_{tt}=-u_i(c^2_1+g_{tt})(h')^2/4$,
we obtain
\begin{eqnarray}
    h=-\frac{4c_1}{u_ic_2}\log(c_4g_{tt}),\   \  e^{-\chi}=\frac{u^4_i}{c^2_1\left[4m^2V(u^2_i,u^4_i) -12\right]}(h')^2g^2_{tt}\, , \label{eq44}
\end{eqnarray}
where $c_4$ is an integration constant.
As $g_{tt}$ decreases near the would-be inner horizon, $h(u)$
exhibits the expected logarithmic growth. The case $c_1=0$ corresponds to the isotropic case, for which the inner horizon survives.

\begin{figure}[htbp]
    \centering
\includegraphics[width=0.55\textwidth]{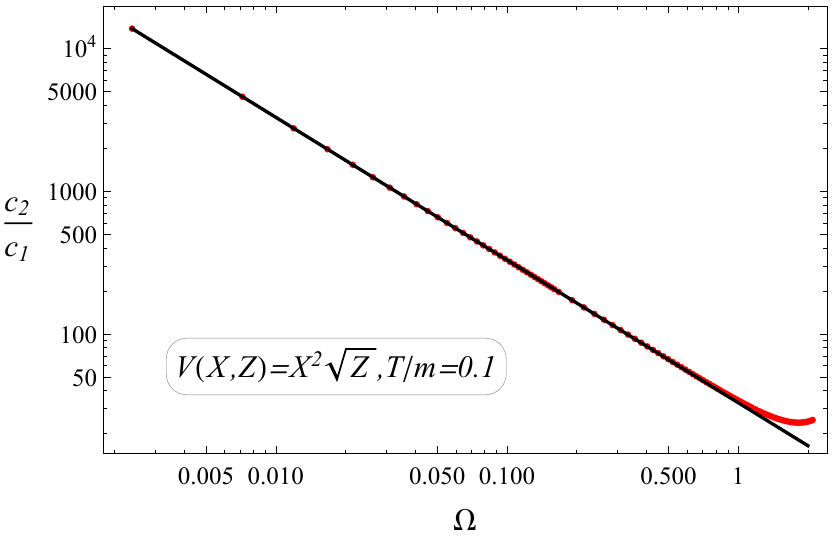}
    \caption{The ratio $c_2/c_1$ as a function of $\Omega$, extracted via $h'=c_2/c_1$ in Table~\ref{tab1} near the would-be inner horizon. As $\Omega$ decreases, $c_2/c_1$ increases, resulting in a more dramatic crossover. Red dots and the black line represent the numerical data and the analytical approximation $c_2/c_1\simeq32.8755/\Omega$, respectively, for the potential $V=X^2\sqrt{Z}$ at $T/m=0.1$.}\label{ratioc}
\end{figure}

The first equation in Eq.~\eqref{eq44} reveals that an increasing $c_2/c_1$ ratio corresponds to decreasing shear deformation.
In fact, ratio $c_2/c_1$ exhibits an inverse proportionality to $\Omega$ in the small shear deformation regime, as shown in Fig.~\ref{ratioc}.
For a sufficiently small shear deformation ($\Omega\ll1$), in the region outside (but in the vicinity of) the would-be inner horizon (i.e., $\delta u<0$), it is evident that $g_{tt}\gg c^2_1/c^2_2$. Therefore, according to Eq.~\eqref{eq43} and the first equation of Eq.~\eqref{eq44}, we find $g_{tt}=\frac{u_ic^2_2}{4}|\delta u|$ and $h'=\frac{4c_1}{u_ic_2}\frac{1}{|\delta u|}$. Similarly, in inside the would-be inner horizon (\emph{i.e.} $\delta u>0$), where $g_{tt}\ll c^2_1/c^2_2$, we obtain $g_{tt}=\text{Exp}\left[-\frac{u_ic^2_2}{4c^2_1}\delta u\right]$ and $h'=c_2/c_1$. We summarize the abrupt changes
in spacetime geometry during the rapid collapse of the ER bridge in Table~\ref{tab1}. The linear decay of $g_{tt}$ near the would-be inner horizon transitions abruptly to an exponential suppression, while $h'(u)$ saturates abruptly from a very small value to a large, finite magnitude under small shear deformation. 
\begin{table}[htbp]
  \centering
\begin{tabular}{c|c}
    \hline\hline  
    \(   u<u_i\, (\delta u<0) \) \hspace{1em} & \hspace{1em} \(   u>u_i\,(\delta u>0)  \)  \\[0.5em]
    \hline
     \(g_{tt}=\frac{u_ic^2_2}{4}|\delta u|\) \hspace{2em} &    \hspace{2.2em} \(g_{tt}=e^{-\frac{u_ic^2_2}{4c^2_1}\delta u}  \)  \\[1em]
     \(  h'=\frac{4c_1}{u_ic_2}\frac{1}{|\delta u|}   \) \hspace{2em} &   \(h'=\frac{c_2}{c_1}\)\\
    \hline\hline  
  \end{tabular}
\caption{The metric component $g_{tt}$ and $h'$ near the would-be inner horizon $u_i$.  For clarity, the shift parameter $c_3$ in Eq.~\eqref{eq43} has been set to zero.} \label{tab1}
\end{table}

Fig.~\ref{ratioc} shows the ratio $c_2/c_1$ as a function of $\Omega$, extracted from the relation $h'=c_2/c_1$ near the would-be inner horizon. The ratio exhibits an inverse proportionality to $\Omega$ in the small-deformation regime. This implies that under small shear deformation, the quantity $(c_2/c_1)^2\delta u$ remains large even for small $\delta u$ near the horizon. Consequently, both the metric component $g_{tt}$ and the anisotropic derivative $h'$ undergo an abrupt change within an extremely narrow interval of the $u$-coordinate, as demonstrated by the numerical results in Fig.~\ref{ERcollapse} and the analytic results in Table~\ref{tab1}. As shown in Fig.~\ref{ERcollapse}\textbf{(a)}, the numerical date exhibit excellent agreement with the preceding analytical analysis.

\subsection{Emergence of domain wall geometry under large shear deformations} \label{sec:scaling}

The nonlinear dynamics associated with the collapse of the ER bridge are suppressed by increasing shear deformation. At sufficiently large shear strain, a geometry characterized by a combined Lifshitz scaling (in both space and time) emerges near the boundary \cite{Baggioli:2020qdg}. This corresponds to a new fixed point on the field theory side that is different from the original $\text{CFT}_4$ UV fixed point. In this work, we further find a domain wall solution interpolating between the geometry with Lifshitz scaling near the UV boundary and that exhibits Kasner scaling behind the horizon.

Since the dynamics occur in a nonlinear regime, the behavior is generally sensitive to the form of the potential. To make further progress, the specific form of 
$V$ must be determined. We consider the benchmark potential $V(X,Z)=X^MZ^N$
with the SSB constraint $M+2N>5/2$. Then, when $\Omega \gg1$, an intermediate geometry emerges between a scale $u_*$ near the UV boundary and another on $u_\text{K}$ inside the black hole. It can be described by an analytic domain wall geometry of the following form:
\begin{eqnarray}
f(u)\simeq f_0\left[1-\left(\frac{u}{u_h}\right)^{3+\frac{9}{\nu^2}}\right],\ \
h(u)\simeq\frac{6}{\nu}\log\left(\frac{u}{u_*}\right),\   \
\chi(u)\simeq\frac{18}{\nu^2}\log\left(\frac{u}{u_h}\right)+\chi_h,   \label{eq48}
\end{eqnarray}
with $\nu=\frac{3M}{M+2N}$ and $f_0=\frac{M \nu^3}{(3+M \nu)(3+\nu^2)}$. $u_*$ satisfies $m^2\left(\frac{e^\Omega}{2}\right)^M u_*^{6M/\nu}=\frac{9}{3+M\nu}$ and $\chi_h\equiv \chi(u_h)=\int^{u_h}_{0}\frac{1}{2}uh'(u)^2du$. Under sufficiently large shear deformation, the analytical solution~\eqref{eq48} accurately describes the bulk geometry from the vicinity of the UV boundary, through the event horizon, and deep into the interior.
Near the UV boundary $u=u_*$, it exhibits an exact scaling symmetry—the anisotropic fixed point identified in Ref.~\cite{Baggioli:2020qdg}. More interestingly, the solution \eqref{eq48} also remains valid in the region behind the event horizon, extending to a scale $u_\text{K}\gg u_h$ that is pushed toward the singularity as the shear deformation increases. At $u=u_\text{K}$, the geometry exhibits a Kasner scaling symmetry, whose features will be detailed in the next section. As will be shown later, 
$u=u_{\text{K}}$ marks the transition point between two Kasner geometries. In the large shear limit $\Omega\rightarrow \infty$, with $u_{\text{K}}$ approaching the singularity, this analytic solution dominates the entire bulk spacetime. 

\begin{figure}[htbp]
    \centering
     \begin{subfigure}[b]{0.54\textwidth}
        \centering
\includegraphics[width=1\textwidth]{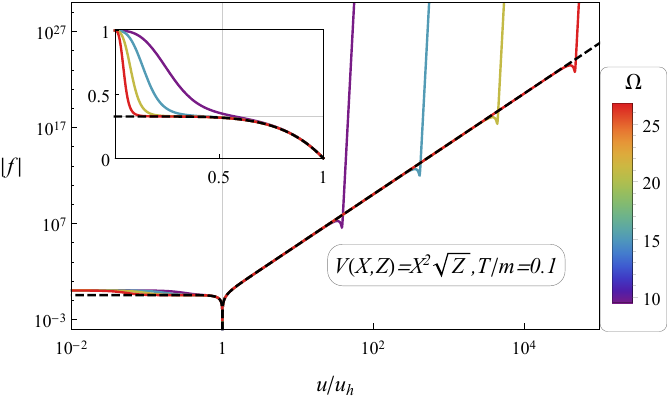}
      \caption{\textbf{(a)}}
    \end{subfigure}
     \hfill \\      
   \begin{subfigure}[b]{0.495\textwidth}
        \centering
\includegraphics[width=1\textwidth]{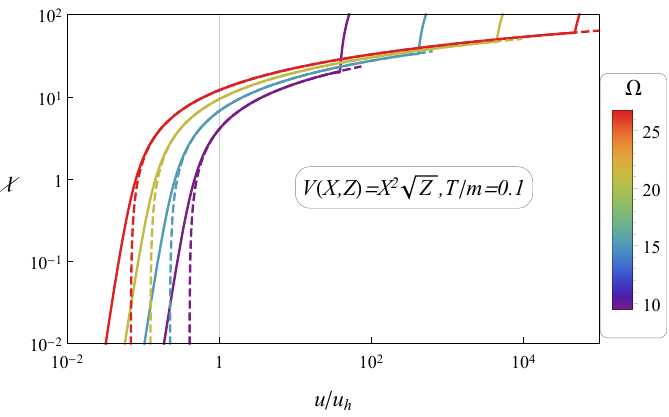}
      \caption{\textbf{(b)}}
    \end{subfigure}
     \hfill       
    \begin{subfigure}[b]{0.495\textwidth}
        \centering
\includegraphics[width=1\textwidth]{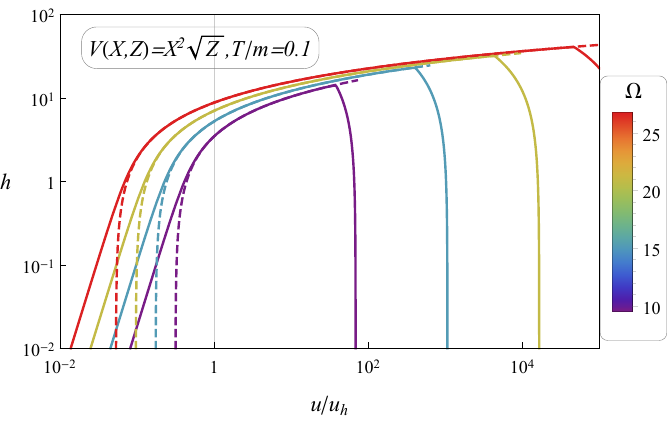}
      \caption{\textbf{(c)}}
    \end{subfigure}
        \caption{Behaviors of $|f(u)|$, $\chi(u)$ and $h(u)$ for different values of $\Omega$. The black dashed line in \textbf{(a)} and the color dashed lines in \textbf{(b)} and \textbf{(c)} represent the analytical solution~\eqref{eq48}. We have fixed $V=X^2\sqrt{Z}$ and $T/m=0.1$.} \label{Nscaling}
\end{figure}

The comparison between the analytic solution and numeric results for the potential $V(X,Z)=X^2\sqrt{Z}$ is presented in Fig.~\ref{Nscaling}. We observe that the analytical solution~\eqref{eq48} closely matches the numerical results in the region between $u_*$ (outside the black hole) and $u_\text{K}$ (inside the black hole).
Appendix \ref{appa} provides more details about the analytical derivation.

\section{Interior dynamics far from the horizon}\label{sec:kasner}

We now turn to the interior dynamics far from the horizon, i.e., the dynamics after the domain wall geometry. While an exact analytical solution is not attainable, we derive self-consistent asymptotic solutions under appropriate approximations.
More precisely, we find that after the collapse of the ER bridge, the spacetime evolves into a regime approaching a Kasner singularity (a Kasner epoch). This Kasner epoch describes a homogeneous but anisotropic spacetime characterized by power-law scaling behavior near the singularity. 
More importantly, our analytical results are strongly supported by the full numerical solutions.

\subsection{Kasner Epoch}\label{sec:KasnerEpoch}
Upon numerical inspection, we find that there is a large-$u$ epoch in which both the axion potential and the cosmological constant become negligible in the equations of motion (the conditions for this approximation will be discussed later). The full equations of motion~\eqref{EOM3}-\eqref{EOM5} reduce to
\begin{eqnarray}
\chi'(u)-\frac{1}{2}uh'(u)^2&=&0\,,\\
\frac{f'(u)}{f(u)}-\frac{3}{u}-\frac{\chi'(u)}{2}&=&0\,,  \\
h''(u)+h'(u)\left[\frac{f'(u)}{f(u)}-\frac{2}{u}-\frac{\chi'(u)}{2}\right]&=&0\,,
\end{eqnarray}
from which we obtain the analytic approximations of the background metric:
\begin{eqnarray}
f(u)=-f_0u^{3+\beta^2},\   \    \chi(u)=2\beta^2\log(u)+\chi_0,\    \  h(u)=2\beta\log(u)+h_0\,,  \label{kasnersolution}
\end{eqnarray}
where $\beta$, $f_0$, $\chi_0$ and $h_0$ are constants. After performing the coordinate transformation $\tau\sim u^{-\frac{3+\beta^2}{2}}$ and diagonalizing the $x-y$ plane leads to a Kasner universe with the metric:
\begin{eqnarray}
    ds^2\simeq -d\tau^2+c_t\tau^{2p_t}dt^2+c_x\tau^{2p_x}d\tilde{x}^2+c_y\tau^{2p_y}d\tilde{y}^2\,, \label{kasnermetric}
\end{eqnarray}
where $c_t$, $c_x$ and $c_y$ are constants. The parameters $p_t$, $p_x$ and $p_y$ are known as the Kasner exponents, which are defined in terms of $\beta$ as follows:
\begin{eqnarray}\label{eqbeta}
    p_t=\frac{-1+\beta^2}{3+\beta^2},\  \  p_x=\frac{2(1-\beta)}{3+\beta^2},\  \ p_y=\frac{2(1+\beta)}{3+\beta^2}\,.  \label{eq56}
\end{eqnarray}
These exponents satisfy the Kasner relations: $p_t+p_x+p_y=1$ and $p^2_t+p^2_x+p^2_y=1$. This illustrates how the holographic flow connects an AdS geometry at the UV boundary to a Kasner universe in the far interior. Nevertheless, the geometry can undergo further changes between different Kasner epochs.

Before proceeding, we examine the interior part of the analytical solution for the domain wall geometry induced by large shear deformation from Section~\ref{sec:scaling}. Substituting Eq.~\eqref{eq48} into the metric ansatz~\eqref{metic}, diagonalizing the $x-y$ directions and considering $u\gg u_h$, we obtain the following geometry:
\begin{eqnarray}\label{IRKasner}
ds^2\simeq u^{1-\frac{9}{\nu^2}}dt^2-u^{-(5+\frac{9}{\nu^2})}du^2+u^{2(\frac{3}{\nu}-1)}d\tilde{x}^2+u^{-2(\frac{3}{\nu}+1)}d\tilde{y}^2\,.
\end{eqnarray}
After the coordinate transformation $\tau\sim u^{-\frac{3+9/\nu^2}{2}}$, the above solution takes precisely the Kasner form of ~\eqref{kasnermetric}. In particular, we can analytically obtain its Kasner exponents~\eqref{eqbeta} with 
\begin{eqnarray}\label{IRbeta}
    \beta=\frac{3}{\nu}=\frac{M+2N}{M}.  \label{eq420}
\end{eqnarray}
It is now clear that for $V=X^M Z^N$, 
a near-horizon Kasner epoch~\eqref{IRKasner} emerges in the large shear deformation limit. This early Kasner epoch can have exponents different from those of the late Kasner epoch in the far interior. The latter develops independently of the shear deformation strength, although the specific values of its Kasner exponents do depend on it.

\begin{figure}[htbp]
      \centering
    \begin{subfigure}[b]{0.52\textwidth}
        \centering
\includegraphics[width=1\textwidth]{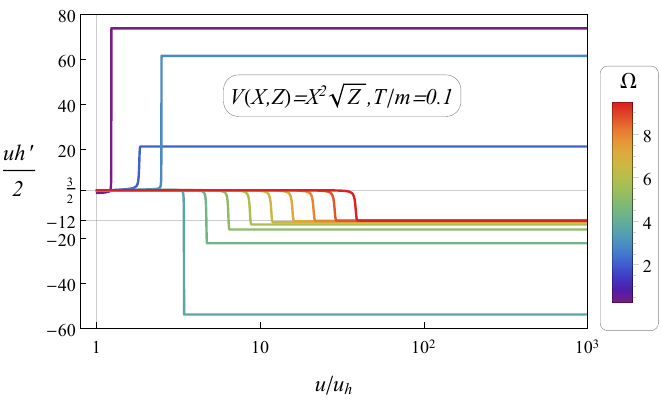}
      \caption{\textbf{(a)}}
    \end{subfigure}
     \hfill   
     \begin{subfigure}[b]{0.47\textwidth}
        \centering
\includegraphics[width=1\textwidth]{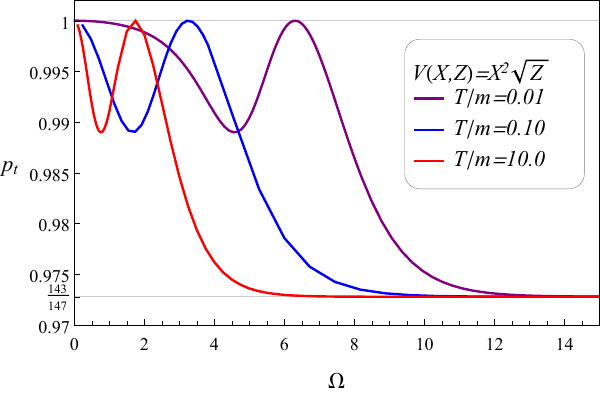}
      \caption{\textbf{(b)}}
    \end{subfigure}
\caption{\textbf{(a)}: The behavior of $\beta=uh'(u)/2$ as a function of $u/u_h$ inside the black hole for different values of $\Omega$.
\textbf{(b)}: The behavior of the Kasner exponent $p_t$ as a function of $\Omega$ at different temperatures. The potential is set to 
$V=X^2\sqrt{Z}$.} \label{figvx2z05}
\end{figure} 
We note from Eq.~\eqref{eqbeta} that $\beta$ plays an important role in determining the Kasner geometry. The value of $\beta$ for a given Kasner universe can be obtained from a plateau of the function $uh'/2$. We show the the variation of $uh'/2$ inside the black hole under different shear deformations in Fig.~\ref{figvx2z05}\textbf{(a)}. We consider the potential $V=X^2\sqrt{Z}$ at the temperature $T/m=0.1$. For each deformation, $uh'/2$ eventually approaches a constant in the far interior, corresponding to a Kasner geometry. Moreover, the Kasner exponent is sensitive to the strength of the shear strain. When the shear
deformation is small, the ER bridge collapses dramatically, resulting in a very large $\beta$ (or equivalently, the Kasner exponent $p_t\rightarrow 1$). As $\Omega$ increases, $\beta$ changes sign from positive to negative. In the large shear strain limit, the exponent $\beta$ approaches a constant. Meanwhile, an intermediate Kasner geometry develops in the region  $u<u_\text{K}$, where $u_\text{K}$ marks the transition between the two Kasner regimes (Kasner geometries near the event horizon and the singularity), corresponding to the emergent domain wall geometry discussed in Section~\ref{sec:scaling}; see also Eq.~\eqref{IRKasner}. The range of this intermediate Kasner epoch grows with increasing $\Omega$. We numerically verify that the alternation between the Kasner epochs is triggered by additional contributions from the axion potential $V$, which are deemed negligible in both Kasner regimes. 

The Kasner exponent $p_t=\frac{-1+\beta^2}{3+\beta^2}$ of the far interior Kasner universe as a function of $\Omega$ is presented in Fig.~\ref{figvx2z05}\textbf{(b)}. For small $\Omega$, it exhibits an oscillatory pattern that depends on the temperature. However, in the large strain limit, $p_t$ approaches a fixed constant uniquely determined by the form of the potential $V$. For the case in Fig.~\ref{figvx2z05}, $p_t=143/147$ as $\Omega\rightarrow\infty$. In the next section, we will summarize more cases of power-law potentials $V(X,Z)=X^{M}Z^N$.

Before ending this section, one should note that the approximate solution~\eqref{kasnersolution} is obtained by neglecting contributions from the axion potential and the cosmological constant, including the terms
\begin{equation}\label{Kterm}
\frac{\Lambda}{f},\quad \frac{V}{f},\quad \frac{\sinh[\Omega-h]V_X}{f}\,.
\end{equation}
To have a stable Kasner universe, we should require all the above terms decay quickly at late interior times ($u/u_h\rightarrow\infty$). From~\eqref{kasnersolution}, one can see that the first term, $\Lambda/f\sim u^{-3-\beta^2}$, can be neglected safely, while the last two terms in~\eqref{Kterm} depend on the form of the axion potential. For a power-law potential $V(X,Z)=X^{M}Z^N$, all the terms in~\eqref{Kterm} behave as power laws in $u/u_h$ under the Kasner solution~\eqref{kasnersolution}. Requiring all these exponents to be negative yields
\begin{eqnarray}\label{condition}
\beta^2-2M|\beta|+3-2M-4N>0\,.  \label{eq510}
\end{eqnarray}
Therefore, for $V(X,Z)=X^{M}Z^N$, once the system enters into a Kasner epoch with an exponent $\beta$ satisfying~\eqref{condition}, it will remain in this epoch all the way to the singularity (see also Appendix \ref{appa}). In contrast, if $\beta$ violates~\eqref{condition}, an alternation to another Kasner epoch is anticipated until the system settles into a stable Kasner universe with a $\beta$ satisfying~\eqref{condition}. This explains the Kasner alternation in Fig.~\ref{figvx2z05} for $V=X^2\sqrt{Z}$ at large shear strain. Our numerical computation confirms that the first Kasner epoch has $\beta_1=3/2$, which agrees quantitatively with our analytical result~\eqref{IRbeta}. It is straightforward to verify that $\beta_1$ violates the stability condition~\eqref{condition}. Thus, as time $``u"$ evolves, the first Kasner epoch is destroyed when contributions from the axion potential become significant to the background geometry at a certain interior time. Consequently, a transition to a second Kasner epoch with exponent $\beta_2=-12$ (or equivalently, $p_t=143/147$) is observed for $\Omega\gg 1$ (see Fig.~\ref{figvx2z05}). The value of $\beta_2$ satisfies the constraint~\eqref{condition}, so the system subsequently settles into a stable Kasner epoch.

\subsection{Interior dynamics for general potentials}\label{sec:othermodel}

Previous discussions have been limited to SSB models with a power-law potential $V(X,Z)=X^M Z^N$  (where $M+2N>5/2$), for which $\Omega$ is directly related to the shear strain. Extending this research to scenarios involving non-spontaneous symmetry breaking and non-polynomial potentials is essential. Nevertheless, in ESB models, $\Omega$ cannot be identified as the shear strain from the dual field theory perspective. We will show that some key results hold for more general axion potentials, while new phenomena can develop for other forms of $V$.

\begin{figure}[htbp]
      \centering
    \begin{subfigure}[b]{0.49\textwidth}
        \centering
\includegraphics[width=1\textwidth]{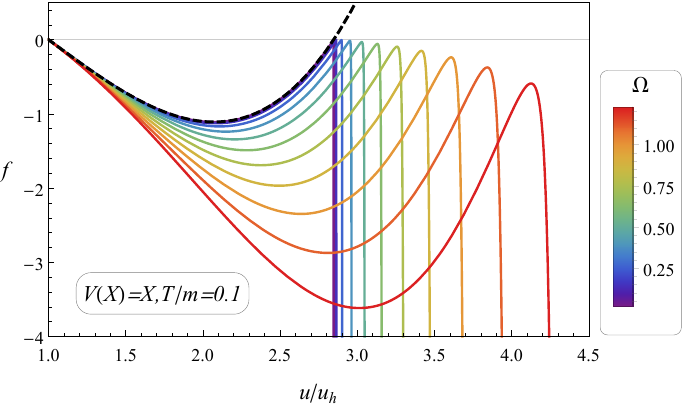}
      \caption{\textbf{(a)}}
    \end{subfigure}
     \hfill   
     \begin{subfigure}[b]{0.49\textwidth}
        \centering
\includegraphics[width=1\textwidth]{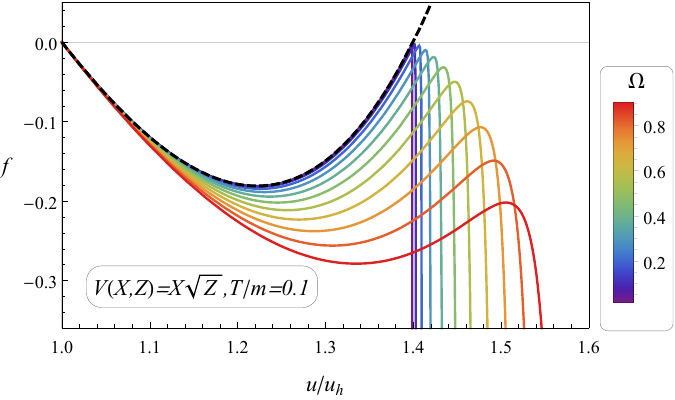}
      \caption{\textbf{(b)}}
    \end{subfigure}
     \hfill \\
     \begin{subfigure}[b]{0.49\textwidth}
        \centering
\includegraphics[width=1\textwidth]{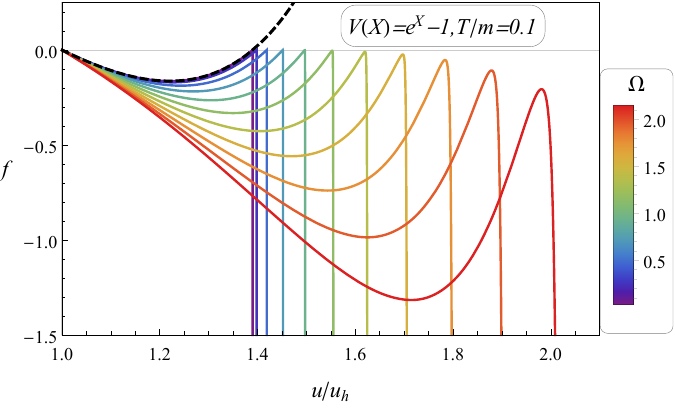}
      \caption{\textbf{(c)}}
    \end{subfigure}
     \hfill   
     \begin{subfigure}[b]{0.49\textwidth}
        \centering
\includegraphics[width=1\textwidth]{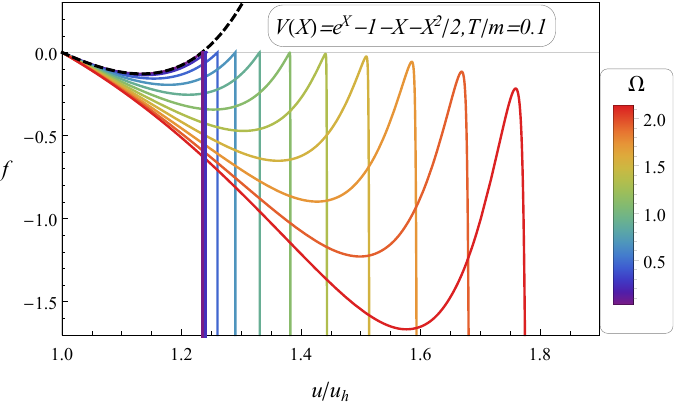}
      \caption{\textbf{(d)}}
    \end{subfigure}
    \caption{The blackening factor $f(u)$ as a function of $u/u_h$ inside the black hole for \textbf{(a)} $V=X$, \textbf{(b)} $V=X\sqrt{Z}$, \textbf{(c)} $V=e^{X}-1$ and \textbf{(d)} $V=e^{X}-1-X-X^2/2$. The dashed curve in each panel shows the solution with an inner horizon in the absence of $\Omega$. The temperature is set to $T/m=0.1$.} \label{othernohrizon}
\end{figure}

As depicted in Fig.~\ref{othernohrizon}, the disappearance of the inner horizon and the occurrence of ER bridge collapse are robust, independent of the choice of potential. This is straightforward to understand from the analysis in Section ~\ref{sec:collapse}, as the potential $V$ becomes negligible near the would-be inner horizon in the small strain limit.

\begin{figure}
  \centering
    \begin{subfigure}[b]{0.52\textwidth}
        \centering
\includegraphics[width=1\textwidth]{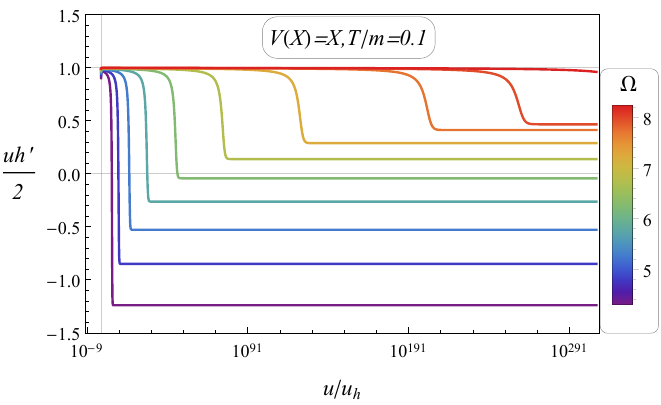}
      \caption{\textbf{(a1)}}
    \end{subfigure}
     \hfill   
     \begin{subfigure}[b]{0.47\textwidth}
        \centering
\includegraphics[width=1\textwidth]{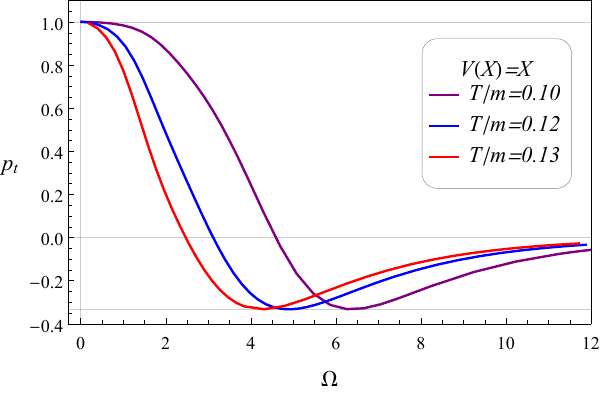}
      \caption{\textbf{(b1)}}
    \end{subfigure}\\
     \begin{subfigure}[b]{0.52\textwidth}
        \centering
\includegraphics[width=1\textwidth]{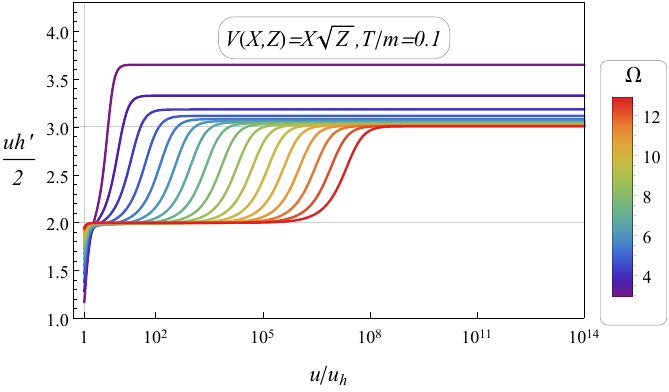}
      \caption{\textbf{(a2)}}
    \end{subfigure}
     \hfill   
     \begin{subfigure}[b]{0.47\textwidth}
        \centering
\includegraphics[width=1\textwidth]{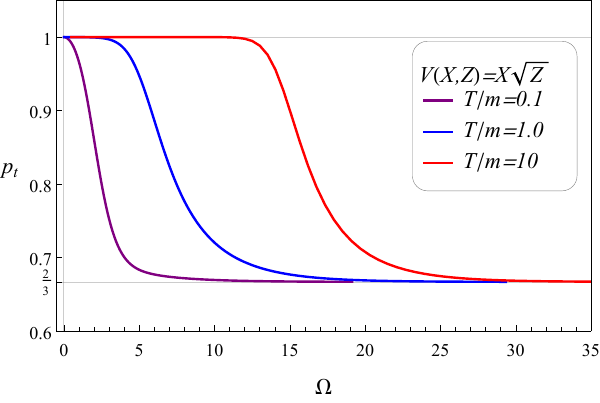}
      \caption{\textbf{(b2)}}
    \end{subfigure}\\   
    
      \begin{subfigure}[b]{0.52\textwidth}
        \centering
\includegraphics[width=1\textwidth]{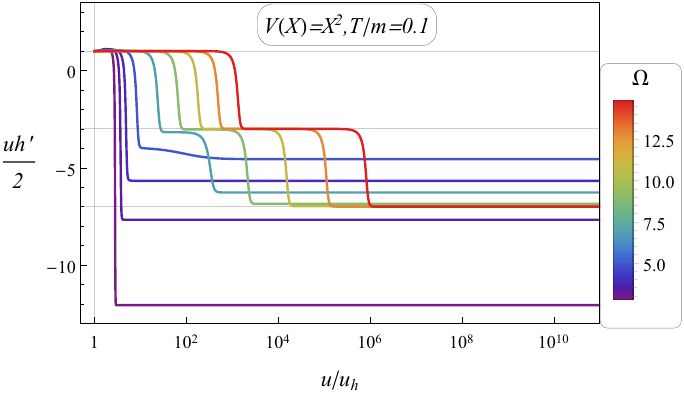}
      \caption{\textbf{(a3)}}
    \end{subfigure}
     \hfill   
     \begin{subfigure}[b]{0.47\textwidth}
        \centering
\includegraphics[width=1\textwidth]{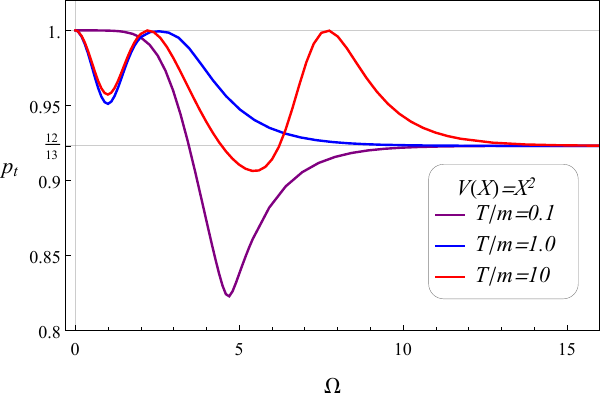}
      \caption{\textbf{(b3)}}
    \end{subfigure}\\   
      \begin{subfigure}[b]{0.52\textwidth}
        \centering
\includegraphics[width=1\textwidth]{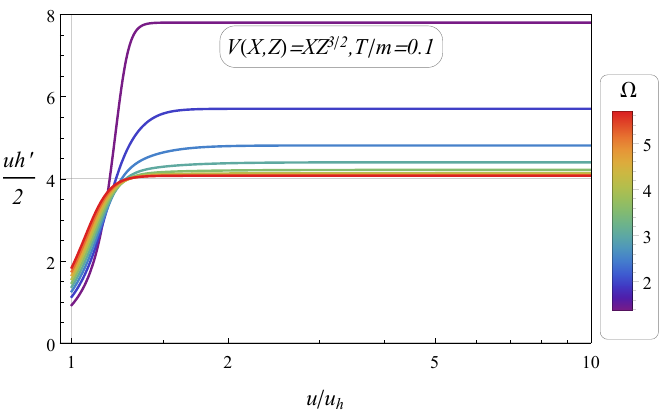}
      \caption{\textbf{(a4)}}
    \end{subfigure}
     \hfill   
     \begin{subfigure}[b]{0.47\textwidth}
        \centering
\includegraphics[width=1\textwidth]{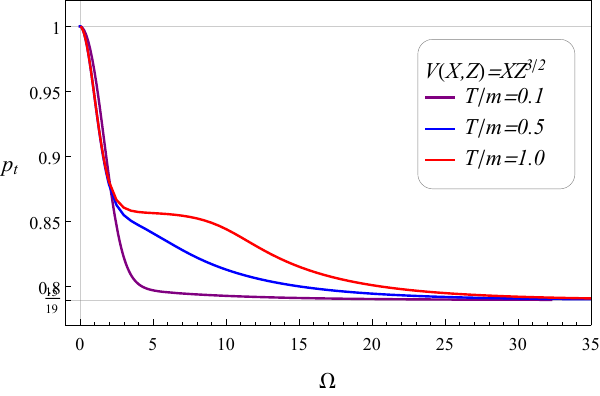}
      \caption{\textbf{(b4)}}
    \end{subfigure}  
\caption{\textbf{(a)}: The function $\beta=uh'/2$ inside the black hole under different values of $\Omega$. \textbf{(b)}: The
Kasner exponent $p_t$ near the singularity as a function of $\Omega$ at three different temperatures. Here we set  $V=X$, $V=X\sqrt{Z}$,  $V=X^2$ and  $V=XZ^{3/2}$, respectively.  } \label{kanergeo}
\end{figure}

For power-law potentials, similar interior dynamics to those in Fig.~\ref{figvx2z05} are found, regardless of whether the translation symmetry is broken spontaneously or explicitly . Representative examples are shown in Fig.~\ref{kanergeo}.  As seen in the left column, before settling into the late-time Kasner epoch, an additional Kasner epoch develops at a finite $u/u_h$ under large shear strain. Its Kasner exponent generally differs from the final Kasner epoch, triggering an alternation between the two epochs, as shown in Figs.~\ref{kanergeo}\textbf{(a1, a2)}. 
In certain cases, more complex behavior arises. For example, with $V=X^2$, as one probes larger $u$, a further Kasner regime develops around $u/u_h=10^4$, see Fig.~\ref{kanergeo}\textbf{(a3)}. Consequently, two Kasner alternations are observed, clearly visible in the red curve at large $\Omega$. Interestingly,  for potentials of the form $V=X Z^N$ with $N>1$, only one Kasner epoch exists after the ER bridge collapse (Fig.~\ref{kanergeo}\textbf{(a4)}). These Kasner alternations can be understood via the stability condition~\eqref{condition} for a power-law potential $V=X^M Z^N$. For $V=X^2$ at large $\Omega$, 
the first two Kasner epochs, characterized by their $\beta$ values, violate the stability condition
~\eqref{condition} and are consequently unstable. In contrast, the third epoch, with $\beta=-7$, satisfies~\eqref{condition}, rendering it stable. Therefore, the system undergoes two Kasner alternations before settling into a stable Kasner universe. For $V=X Z^N$, the exponent $\beta=1+2N$ from the early Kasner epoch~\eqref{IRKasner} satisfies
\begin{eqnarray}
\beta^2-2M|\beta|+3-2M-4N=4N(N-1)>0\, ,\   \  (M=1)\,.
\end{eqnarray}
Therefore, the early Kasner epoch is stable for $N>1$, and no Kasner alternation is anticipated, as confirmed in Fig.~\ref{kanergeo}\textbf{(a4)}. In contrast, the early Kasner epoch is unstable for $N<1$, resulting in at least one Kasner alternation, as seen in Fig.~\ref{kanergeo}\textbf{(a1, a2)}.

The corresponding Kasner exponent $p_t$ for the far-interior epoch versus $\Omega$ is shown in the right column of Fig.~\ref{kanergeo}. Its behavior is sensitive to both the form of the potential and the temperature. For example, the oscillatory pattern disappears for $V=X\sqrt{Z}$ and $V=X Z^{3/2}$. The value of $p_t$ can also be negative, as shown in the linear axion case $V=X$. However, in the large $\Omega$ limit, $p_t$ approaches a constant independent of temperature. We summarize the Kasner exponents for the far-interior geometry of various power-law potentials $V=X^N Z^M$ in Table.~\ref{tab:Kasner}. The values of $p_t$ in the last column are obtained by numerically solving the full set of Eq.~\eqref{EOM3}-\eqref{EOM5}. One can verify that they all satisfy condition~\eqref{condition}. 

\begin{table}[htbp]  
  \centering
  \begin{tabular}{c|c|c|c}
    \hline\hline  
    \(M\) & \(N\) & \(\beta(\Omega\rightarrow\infty)\) & \(p_t(\Omega\rightarrow\infty)\) \\[0.2em]
     \hline  
    1 & 0 &  1    & 0  \\
    1 &  0\(<N\leq\)1 &   3  &  2/3 \\
    1 & \(N\geq1\)  &   \(\frac{M+2N}{M} \) &  \(\frac{-M^2+(M+2N)^2}{3M^2+(M+2N)^2} \)  \\
     2 &  0  &  -7 &  12/13  \\
     2 &  1/4  & -6 &  35/39 \\
     2 &  1/2  & -12 &  143/147 \\
     2 &  3/4  & -30 &  899/903  \\
     3 &  1  & -9 &  20/21  \\
   \hline\hline   
  \end{tabular}
  \caption{Kasner exponents for different values of \(M\) and \(N\) of power-law potentials $V(X,Z)=X^MZ^N$ under large anisotropy strength. Note that $\beta$ is obtained through high-precision numerical extraction, while $p_{t}$ is calculated using Eq.~\eqref{eq56}.}\label{tab:Kasner}
\end{table}

\begin{figure}
  \centering
\includegraphics[width=0.8\textwidth]{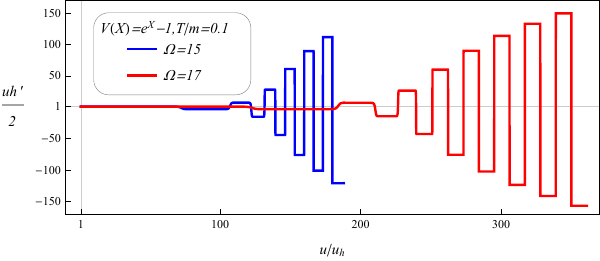}
\caption{The function $\beta=uh'(u)/2$ insider the black hole under two different values of $\Omega$ for the potential $V(X)=e^X-1$ at the temperature $T/m=0.1$.}\label{VexpX}
\end{figure}

Understanding other cases is considerably more challenging due to the highly nonlinear nature of the interior dynamics. We find that, for exponential-like potentials, the solution also enters a regime described by anisotropic Kasner universe at later interior time. However, instead of settling into to a stable Kasner universe, an endless alternation of Kasner epochs occurs toward the singularity. In Fig.~\ref{VexpX}, we illustrate the interior dynamics for $V=e^{X}-1$. As the deep interior is approached, the dynamics becomes dominated by the leading exponential behavior $V\sim e^X$ and the oscillation amplitude continues to grow~\footnote{However, the behavior extremely close to the singularity remains elusive, largely due to the limitations of our computational resources.}. This behavior emerges irrespective of the value of $\Omega$. Similar features have been observed in the isotropic black holes with the scalar hair, induced by a symmetric super-exponential scalar potential~\cite{Cai:2020wrp,Hartnoll:2022rdv}. Moreover, we can show that for a potential $V(X)$ diverging exponentially or faster, its contribution cannot be neglected, thus precluding any stable Kasner epoch. Suppose that the geometry is in a Kasner regime characterized by $h\sim 2\beta\ln u$ and $f\sim u^{3+\beta^2}$ at large $u$. For $V\sim e^X$, we have
\begin{eqnarray}
\frac{V}{f}\sim \frac{e^{u^2\cosh(\Omega-h)}}{f}\sim \frac{e^{u^{2+2|\beta|}}}{u^{3+\beta^2}}\,,\\
\frac{\sinh(\Omega-h)V_X}{f}\sim \frac{\sinh(\Omega-h)e^{u^2\cosh(\Omega-h)-1}}{f}\sim \frac{e^{u^{2+2|\beta|}}}{u^{3-2|\beta|+\beta^2}}\,.
\end{eqnarray}
Clearly, in both terms the numerator diverges much faster than the denominator. Therefore, they must become significant as $u$ increases, leading to a strong deviation from the Kasner solution. 

A pretty simple case is that of a potential bounded from above. From Eq.~\eqref{eq211}, an isotropic black hole has no inner horizon if $V<3/m^2$. Consequently, there is no instability triggered by shear deformation, unlike in previous cases. As a result, both the collapse of the ER bridge and the abrupt change in spatial anisotropy vanish. This demonstrates that the intricate dynamics are linked to the instability of the would-be inner horizon induced by shear deformation. Furthermore, since the potential has a finite upper bound, the stability condition~\eqref{condition} is readily satisfied. We therefore obtain a stable Kasner epoch all the way to the spacelike singularity.

To close this section, we draw a comparison between the Kasner transition in our model and that predicted by the standard BKL dynamics with pure gravitational walls. In the latter scenario, the exponents of the subsequent Kasner epoch are uniquely fixed by those of the preceding epoch via the relation (3.14) derived in~\cite{Belinsky:1970ew}. In our model, the exponents of the Kasner epoch induced by large shear deformation depend exclusively on potential form. If the transition were to adhere to the usual BKL dynamics of~\cite{Belinsky:1970ew}, the subsequent Kasner epoch would be fixed and unique. However, as is clearly illustrated in  the right panels of Fig.~\ref{figvx2z05} and Fig.~\ref{kanergeo}), the subsequent Kasner epoch 
is not fixed. Instead, it depends sensitively on both the magnitude of the shear strain and the temperature. We therefore conclude that our results go beyond the transition rule of the standard BKL dynamics.

\section{Discussion and outlook}\label{sec:conclusion}

In this work, we explore the internal structure of black holes under finite shear strain in a class of holographic systems with broken spatial symmetries, which have been used to study nonlinear elastic responses of solids to mechanical deformations, including stress-strain curves analogous to those of real materials as well as the mechanism of the elastic failure. It is found that the interior dynamics of black holes, significantly affected by the axion fields (which induce the spatial anisotropy), exhibits rich and intriguing behaviors.

For the general potential of the axions that satisfies the theoretical self-consistency and allows the existence of the inner horizon in the isotropic case, we demonstrate that the shear deformation always removes the inner horizon of the black hole, resulting in a spacelike singularity, thereby preserving the strong cosmic censorship conjecture. Meanwhile, the ER bridge undergoes a rapid collapse when the shear deformation is small. More precisely, the collapse becomes more violent for smaller shear deformations. In contrast to previous cases where the collapse is driven by homogeneous (charged) scalar or vector hair ~\cite{Hartnoll:2020rwq,Hartnoll:2020fhc,Cai:2020wrp,Cai:2021obq}, these features are induced solely by the shear anisotropy in the present model. Nevertheless, the ER collapse will eventually disappear 
when the shear is sufficiently enhanced. In particular, for a power-law potential $V(X,Z)=X^MZ^N$, this process is superseded by the emergence of an intermediate domain-wall geometry, which flows from the UV anisotropic Lifshitz fixed point toward a Kasner epoch. Eventually, the geometry either tends to stable Kasner universes, or exhibit an endless alternation of different Kasner epochs toward the singularity, depending on the form of the potential.

While several universal features of the interior dynamics have been identified, many open problems remain. We have numerically observed alternations between different Kasner epochs, but a comprehensive theoretical understanding of this dynamics is still lacking. In particular, the case with an exponential axion potential exhibits very rich internal behaviors that are far from being understood (see Fig.~\ref{VexpX}). 
Moreover, it would also be interesting to investigate the holographic correspondences of these internal dynamics in the dual field theory, particularly their implications for correlation functions, entanglement entropy, quantum complexity and quantum chaos. (see, \emph{e.g.},~\cite{Mansoori:2021wxf,Xu:2023fad,Caceres:2024edr,An:2022lvo,Prihadi:2025czn}). Our current work has been limited to a class of simple holographic systems. Nevertheless, one anticipates even richer interior dynamics may appear when the present model is coupled with some other fields (\textit{e.g.} vector field or dilaton) \cite{Gouteraux:2016wxj,Baggioli:2016oqk,Amoretti:2017axe}. We leave these subjects for future work. 

\acknowledgments
Yuanceng Xu would like to thank Matteo Baggioli for sharing his experience in numerical calculations. This work is supported by the National Natural Science Foundation of China (NSFC) under Grants No.\,12275038, No.\,12447178, No.\,12525503 and No.\,12447101.

\appendix
\section{On the constraint $V_X\geq0$} \label{appa1}
In this section, we show that the non-negativity of $V_X$ should be satisfied both from  theoretical self-consistency of the bulk theory as well as the requirement for a well-defined elastic property of the boundary system.

Note that our model is the Stückelburg formalism of the Lorentz-violating massive gravity dual to broken translations on boundary. The effective mass of the spin-2 graviton can be determined in the following way \cite{Hartnoll:2016tri}: 
\begin{eqnarray}
m(u)^2=g^{xx}T_{xx}-\frac{\delta T_{xy}}{\delta g_{xy}}=2m^2\bar{V}_{X}u^2,
\end{eqnarray}
where $\bar{V}_X\equiv d V/dX|_{X=\bar{X},Z=\bar{Z}}$. Then, to avoid the tachyonic instability, we should require that $V_X\geq0$. 

Furthermore, due to the Lorentz-violating mass terms, the vector and tensor modes can be free from the Boulware-Deser ghost \cite{Rubakov:2004eb,Dubovsky:2004sg}. Nevertheless, there still could be ghost instability from the scalar sector. Assuming isotropic backgrounds, in the decoupling limit (valid for small $m^2$), we perturb the scalar fields $\phi^I=\bar{\phi}^I+\delta \phi^I$ and expand $V$ to quadratic order, yielding:
\begin{eqnarray} \label{} 
V_{(2)}\sim \left(\bar{X}\bar{V}_X+2\bar{Z}\bar{V}_Z\right)\partial_\mu\delta\phi^I\partial^\mu\delta\phi^I-2\bar{Z}\bar{V}_Z\left(\varepsilon^{iI} \partial_i\delta\phi^I\right)^2\nonumber\\ 
+\left(\bar{X}^2\bar{V}_{XX}+4\bar{X}\bar{Z}\bar{V}_{XZ}+4\bar{Z}^2\bar{V}_{ZZ}\right)\left(\delta_I^i\partial_i\delta\phi^I\right)^2,
\end{eqnarray}
where subscripts `$X$' and `$Z$' denote partial derivatives with respect to $X$ and $Z$, respectively. To avoid ghost instabilities and gradient instabilities, we must impose the following two constraints:
\begin{eqnarray}   
\bar{X}\bar{V}_X+2\bar{Z}\bar{V}_Z\geq0,\   \  
\end{eqnarray}
and
\begin{eqnarray} 
\bar{V}_X\geq 0,\   \ \bar{X}^2\bar{V}_{XX}+4\bar{X}\bar{Z}\bar{V}_{XZ}+4\bar{Z}^2\bar{V}_{ZZ}\geq -\bar{X}V_X-2\bar{Z}\bar{V}_Z.
\end{eqnarray}
Note that the second constraint (which gives rise to a positive square of the transverse sound's speed) above also ensures a positive define mass squared of the graviton. In addition, we  require $V(\bar{X}\rightarrow0,\bar{Z}\rightarrow0)\rightarrow 0$ so that the theory admits asymptotic AdS  solutions. 

Next, we examine the physical interpretation of the scalar background profile $\bar{\phi}^I(x^i)\sim x^i$ and how the form of the potential $V$ governs the mechanism that breaks translations at the boundary. Approaching the boundary, the gravity sector and matter fields become increasingly decoupled from each other. In this regime, the linearized equation for $\delta \phi^I$ is given by
\begin{eqnarray}\label{UVeq}
\partial_\mu \left[\sqrt{-\bar{g}} \left( \bar{V}_X \partial^\mu \delta \phi^I + \bar{V}_Z \bar{Z} \bar{\mathcal{I}}_{IJ} \partial^\mu \delta \phi^J \right) \right] = 0,
\end{eqnarray}
where $\mathcal{I}_{IJ}$ is the inverse of $\mathcal{I}^{IJ}$ defined in Eq.~\eqref{eqIJ}. Suppose that $V(\bar{X},\bar{Z})\sim u^A$ with the exponent $A>0$ near the boundary. Substituting this scaling into Eq. (\ref{UVeq}), the scalar fluctuations exhibit the near-boundary behavior
\begin{eqnarray} \label{UVdeltaphi}  
\delta \phi^I(u,t,x^I)\approx\delta \phi^I_{(0)}(t,x^I)+\dots+\delta\phi^I_{(1)}(t,x^I)\,u^{5-A}+\dots,
\end{eqnarray}
where we have used $\bar{X}\sim u^2$ and $\bar{Z}\sim u^4$. For $A>5$, the near-boundary expansion of the field $\phi^I$ can then be organized in the following order:
\begin{eqnarray}   
\phi^I(u,t,x^i)
\approx \left(0+\delta \phi_{(1)}^I(t,x^i)\right)\, u^{5-A}+\dots+\left(\bar{\phi}_{(0)}^I(x^i)+\delta \phi_{(0)}^I(t,x^i)\right)+\mathcal{O}(u).
\end{eqnarray}
Evidently, the background profile $\bar{\phi}^I=\bar{\phi}_{(0)}^I(x^i)\sim x^i$ enters the subleading term in this expansion. It should be interpreted as a non-zero vacuum expectation value, $\langle \Phi^I\rangle\sim x^i$, related to a phase with SSB (in the standard quantization scheme). For the benchmark potential $V(X,Z)=X^MZ^N$, the condition $A>5$ requires that $M+2N>5/2$ \cite{Alberte:2017oqx}. When $0<A<5$, the $u-$independent term dominates the UV expansion (\ref{UVdeltaphi}), and the profile $\bar{\phi}^I=\bar{\phi}_{(0)}^I(x^i)\sim x^i$ is dual to the external source of the $\Phi^I$ operator that explicitly breaks translations.

We now turn to the requirement for the well-defined elastic property in the boundary
system to derive the basic constraints on $V_X$ necessary for the consistency of the general model~\eqref{action}. Treating $m^2$ as a small
parameter again, the shear stress at first order is given by ~\cite{Baggioli:2020qdg}
\begin{equation}
 \sigma=\frac{1}{2} m^2\epsilon\sqrt{4+\epsilon^2}  \int_0^{u_h} \frac{V_X((1+\epsilon^2/2)s^2, s^4)}{s^2} d s\,.
\end{equation}
We can analytically derive a simple expression for the linear shear modulus:
\begin{equation}
\mathcal{G}=\frac{d\sigma}{d\epsilon}\Big|_{\epsilon=0}=m^2  \int_0^{u_h} \frac{V_X(s^2, s^4)}{s^2} d s\,.  
\end{equation}
Given that the strain and $u_h$ can vary arbitrarily, a positive shear modulus is guaranteed by the condition $V_X\geq0$. For the benchmark potential $V(X,Z)=X^MZ^N$, the above requirements imply $M\geq0$ together with the condition for the SSB scenario that is 
$M+2N>5/2$.

\section{Analytic analysis of domain wall geometry}\label{appa}

In this section, we present an analytical verification about the domain wall geometry~\eqref{eq48} induced by large shear deformation, as discussed in Sec.~\ref{sec:scaling}.
We consider the following benchmark potential
\begin{eqnarray}
V(X,Z)=X^MZ^N.
\end{eqnarray}
From the equations of motion~\eqref{EOM4}-\eqref{EOM5} and based on extensive numerical evidence, we infer semi-analytically that $h(u)$, in the limit of large shear deformation, admits a logarithmic form
\begin{eqnarray}
h(u)=\frac{6}{\nu} \log \left(\frac{u}{u_*}\right), \label{eq422}
\end{eqnarray}
where $\nu=\frac{3M}{M+2N}$ and $u_*$ is an emergent scale that will be fixed later. 
Applying Eq.~\eqref{eq422}, the background equations become the following forms:
\begin{eqnarray}
\left[u^{-(3+9/\nu^2)}f\right]'&=&u^{-(4+9/\nu^2)}\left[m^2V(\bar{X},\bar{Z})-3\right],\nonumber\\
&=&u^{-(4+9/\nu^2)}\left\{m^22^{-M}\left[e^\Omega\left(\frac{u}{u_*}\right)^{-6/\nu}+e^{-\Omega}\left(\frac{u}{u_*}\right)^{6/\nu}  \right]^{M}u^{6M/\nu}-3  \right\},\nonumber\\
\label{eq423}\\
\left[u^{-(3+9/\nu^2)}f\right]'&=&u^{-(4+9/\nu^2)}\frac{m^2\nu}{3} V_h(\bar{X},\bar{Z}),\nonumber\\  
&=&-u^{-(4+9/\nu^2)}\frac{M\nu}{3}m^22^{-M}\left[e^\Omega\left(\frac{u}{u_*}\right)^{-6/\nu}-e^{-\Omega}\left(\frac{u}{u_*}\right)^{6/\nu}  \right]\nonumber\\
&&\left[e^\Omega\left(\frac{u}{u_*}\right)^{-6/\nu}+e^{-\Omega}\left(\frac{u}{u_*}\right)^{6/\nu}  \right]^{M-1}u^{6M/\nu},
\label{eq424}
\end{eqnarray}
where $V_h(\bar{X},\bar{Z})\equiv \partial V/\partial h|_{X=\bar{X},Z=\bar{Z}}=-M\sinh{[\Omega-h(u)]}\cosh{[\Omega-h(u)]}^{M-1}u^{6M/\nu}$. Although the expressions on the right-hand side of the two equations above seem different, they are approximately equal within the range of interest (except for a very narrow region inside the event horizon). Therefore, the solution
obtained from the two equations differ only slightly.
We consider sufficiently large shear $(\Omega\gg1)$ and the region where $e^\Omega\left(u/u_*\right)^{-6/\nu}\gg e^{-\Omega}\left(u/u_*\right)^{6/\nu}$ is satisfied. Then, Eqs.~\eqref{eq423} and~\eqref{eq424} reduce to:
\begin{eqnarray}
\left[u^{-(3+9/\nu^2)}f\right]'&\simeq& u^{-(4+9/\nu^2)}\left[m^22^{-M}e^{M\Omega}u_*^{6M/\nu}-3  \right],\label{eq425}\\
\left[u^{-(3+9/\nu^2)}f\right]'&\simeq&-u^{-(4+9/\nu^2)}\frac{M\nu}{3}m^22^{-M}e^{M\Omega}u_*^{6M/\nu}.\label{eq426}
\end{eqnarray}
By equating the right-hand sides of the two equations, the constant $u_*$ can be fixed via the relation
\begin{eqnarray}
m^2\left(\frac{e^\Omega}{2}\right)^M u_*^{6M/\nu}&=&\frac{9}{3+M\nu}.\label{eq429}
\end{eqnarray}
Obviously, when the limit $\Omega\rightarrow \infty$ is taken, we have that $u_*\rightarrow 0$.  
Using the form of $u_*$, Eqs.~\eqref{eq425} and~\eqref{eq426} become as follows:
\begin{eqnarray}
\left[u^{-(3+9/\nu^2)}f\right]'\simeq-\frac{3M\nu}{3+M\nu}u^{-(4+9/\nu^2)}.   \label{eqB9}
\end{eqnarray}
By integrating Eq.~\eqref{eqB9} from $u_h$ to $u$, we can easily obtain the following solution:
\begin{eqnarray}
f(u)=\frac{M \nu^3}{(3+M \nu)(3+\nu^2)}\left[1-\left(\frac{u}{u_h}\right)^{3+\frac{9}{\nu^2}} \right]=f_0\left[1-\left(\frac{u}{u_h}\right)^{3+\frac{9}{\nu^2}} \right]. \label{eq434}
\end{eqnarray}
Considering Eq.~\eqref{EOM3} and setting $\chi(u_h)=\chi_h$, we achieve that
\begin{eqnarray}
\chi(u)=\frac{18}{\nu^2}\log\left(\frac{u}{u_h}\right)+\chi_h. \label{eq434a}
\end{eqnarray}
This is exactly the analytic solution we mentioned in the main text. 

One can estimate the range of the region where \eqref{eq422}, \eqref{eq425} and \eqref{eq426} become good approximations. Setting $e^\Omega\left(u/u_*\right)^{-6/\nu}\simeq e^{-\Omega}\left(u/u_*\right)^{6/\nu}$, the (possible) upper bound on the range can be fixed as 
\begin{eqnarray}
u_0\equiv 2^{-\frac{6}{\nu}}\left[\frac{9}{m^2(3+M \nu)}\right]^{-\frac{6}{M\nu}}.
\end{eqnarray}
Together with the requirement of the non-negativity of $h(u)$, we conclude that the analytic solution is a very good approximation at least in the window that $u_*\ll u\ll u_0$.

\begin{figure}[hbtp!]
  \centering
\includegraphics[width=0.5\textwidth]{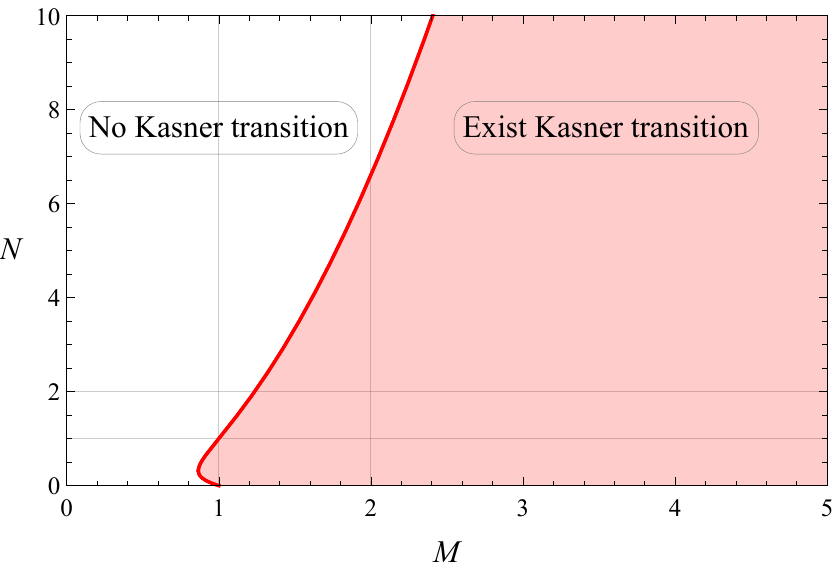}
\caption{For a power-law potential $V(X,Z)=X^MZ^N$, whether a Kasner transition occurs in the large shear limit is determined by the exponents $M$ and $N$.  }\label{nokasner}
\end{figure}

Now, we turn to consider another region where
$u\gg u_0$ such that $e^\Omega\left(u/u_*\right)^{-6/\nu}\ll e^{-\Omega}\left(u/u_*\right)^{6/\nu}$ even for the large value of $\Omega$. In this region, Eqs.~\eqref{eq423} and~\eqref{eq424} reduce to 
\begin{eqnarray}
\left[u^{-(3+9/\nu^2)}f\right]'&\simeq& u^{-(4+9/\nu^2)}\left[m^22^{-M}e^{-M\Omega}\left(\frac{u}{u_*}\right)^{6M/\nu}u^{6M/\nu}-3  \right],\nonumber\\
&\simeq&u^{-(4+9/\nu^2)}\left[m^{4}2^{-2M}\frac{(3+M\nu)}{9}u^{\frac{12M}{\nu}}-3\right],\label{eq4292}\\
\left[u^{-(3+9/\nu^2)}f\right]'&\simeq&-u^{-(4+9/\nu^2)}\frac{M\nu}{3}m^22^{-M}\left[-e^{-\Omega}\left(\frac{u}{u_*}\right)^{6/\nu}\right] \left[e^{-\Omega}\left(\frac{u}{u_*}\right)^{6/\nu}\right]^{M-1}u^{6M/\nu},\nonumber\\
&\simeq&u^{-(4+9/\nu^2)}\frac{M\nu}{3}m^42^{-2M}\frac{(3+M\nu)}{9}u^{\frac{12M}{\nu}}.\label{eq4301}
\end{eqnarray}
By integrating Eqs.~\eqref{eq4292} and~\eqref{eq4301} from $u_0$ to $u$ and considering the boundary condition $f(u_0)\simeq-f_0\left(\frac{u_0}{u_h}\right)^{3+\frac{9}{\nu^2}}$, we obtain the following solutions
\begin{eqnarray}
f(u)&=&f(u_0)\left(\frac{u}{u_0}\right)^{3+\frac{9}{\nu^2}}+u^{3+\frac{9}{\nu^2}}\int^{u}_{u_0}u^{-(4+9/\nu^2)}\left[m^{4}2^{-2M}\frac{(3+M\nu)}{9}u^{\frac{12M}{\nu}}-3\right]du,\nonumber\\
&\simeq&-f_0\left(\frac{u}{u_h}\right)^{3+\frac{9}{\nu^2}}+\mathcal{F}(u)+\frac{\nu^2\left[1-\left(\frac{u}{u_0}\right)^{3+\frac{9}{\nu^2}}\right]}{3+\nu^2},\label{eq431}\\
f(u)&=&f(u_0)\left(\frac{u}{u_0}\right)^{3+\frac{9}{\nu^2}}+u^{3+\frac{9}{\nu^2}}\int^{u}_{u_0}
u^{-(4+9/\nu^2)}\frac{M\nu}{3}m^42^{-2M}\frac{(3+M\nu)}{9}u^{\frac{12M}{\nu}}du,\nonumber\\
&\simeq&-f_0\left(\frac{u}{u_h}\right)^{3+\frac{9}{\nu^2}}+\frac{M\nu}{3}\mathcal{F}(u),\label{eq432}
\end{eqnarray}
where
\begin{eqnarray}
\mathcal{F}(u)&=&\frac{m^{4}2^{-2M}(3+M\nu)\nu^2}{27(4M\nu-3-\nu^2)}\left[\left(\frac{u}{u_0} \right)^{\frac{12M}{\nu}}   -\left(\frac{u}{u_0} \right)^{3+\frac{9}{\nu^2}} \right]u^{\frac{12M}{\nu}}_0\,,\nonumber\\
&=&\frac{3\nu^2\left[\left(\frac{u}{u_0} \right)^{\frac{12M}{\nu}}-\left(\frac{u}{u_0} \right)^{3+\frac{9}{\nu^2}} \right]}{(4M\nu-3-\nu^2)(3+M\nu)}\,.
\label{eq432extra}
\end{eqnarray}
Then, the solution \eqref{eq422} can be a good approximation only if \eqref{eq431} and \eqref{eq432} are consistent with each other. Due to the fact that $u_h\ll u_0$\footnote{This can be seen via solving Eq.\eqref{EOM4} near the horizon in the large shear limit, which gives $u_h\sim e^{-\frac{3\nu\,\Omega}{2(9+\nu^2)}}\rightarrow 0,$ as $\Omega\rightarrow \infty$ \cite{Baggioli:2020qdg,Pan:2021cux}.}, when  $4M\nu-3-\nu^2>0$, the blackening factor can be expressed as  
\begin{eqnarray}
f(u)\simeq-f_0\left(\frac{u}{u_h}\right)^{3+\frac{9}{\nu^2}}, \label{eq4321}
\end{eqnarray}
 matching the leading
 behavior of \eqref{eq434} in this region, which implies that the validity region of the analytic expressions can be extended up to a new scale $u_\text{K}\gg u_0$ in the deep interior of the black hole. Indeed, the numeric result shown in Fig.~\ref{Nscaling} is in good agreement with the analytic expressions \eqref{eq48}, ranging from the scale $u_*$ close to the AdS boundary up to the scale $u_\text{K}$ inside the black hole, where $u_\text{K}$ can be estimated as follows:
\begin{eqnarray}
\left(\frac{u_\text{K}}{u_h}\right)^{3+\frac{9}{\nu^2}}\simeq\left(\frac{u_\text{K}}{u_0}\right)^{\frac{12M}{\nu}}\,,
\end{eqnarray}
for which the function $f(u)$ solved under the approximation~\eqref{eq422} as well as large $\Omega$ limit approaches zero around $u_\text{K}$, \emph{i.e.} the development of an inner horizon. This is impossible as the inner Cauchy horizon is removed completely by the shear anisotropy as demonstrated in Section~\ref{sec:nohorizon}.
From the equation above, we fix that
\begin{eqnarray}
u_\text{K}&\simeq&\left[u^{3+\frac{9}{\nu^2}}_h u^{-\frac{12M}{\nu}}_0\right]^{\frac{1}{3+\frac{9}{\nu^2}-\frac{12M}{\nu}}}\nonumber\\
&\sim& e^{\frac{3\nu(3+\nu^2)\,\Omega}{2(9+\nu^2)(4M\nu-3-\nu^2)}}. \   \   
\end{eqnarray}
On the contrary, in the case of $4M\nu-3-\nu^2\leq0$, the emergent domain wall geometries at large shear deformations and the Kasner geometry near the singularity share an identical solution. In other words, there is no transition of the Kasner geometry. In fact, we find that $-\frac{3}{\nu^2}(4M\nu-3-\nu^2)=\beta^2-2M|\beta|+3-2M-4N>0$ with $\beta=(M+2N)/M$ is exactly the condition~\eqref{eq510} for geometries with no Kasner transition. An illustration of the parameter space for the power-law potential, where 
the Kasner transition exists or does not exist, is shown in Fig. \ref{nokasner}.

\bibliographystyle{JHEP}
\bibliography{biblio.bib}

\end{document}